\documentclass[12pt]{iopart}
\usepackage{cite,amssymb,amsfonts,graphicx,bm,dcolumn,mathrsfs}

\usepackage{epsfig}

\newcommand{\dd}{\mbox{d}}

\newcommand{\er}{\mathrm{e}}
\newcommand{\bea}{\begin{eqnarray}}
\newcommand{\eea}{\end{eqnarray}}
\newcommand{\be}{\begin{equation}}
\newcommand{\ee}{\end{equation}}

\begin{document}

\title[Three-state Potts model with additional mean-field interaction]{Phase diagram
of the one-dimensional three-state Potts model with an additional mean-field interaction}
\author{Alessandro Campa$^{1,2}$\footnote{Corresponding author}, Vahan Hovhannisyan$^3$, Stefano Ruffo$^{4,5,6}$ and Andrea Trombettoni$^{7,8}$}
\address{$^1$ National Center for Radiation Protection and
Computational Physics, Istituto Superiore di Sanit\`{a}, Viale Regina Elena 299, 00161 Roma, Italy}
\address{$^2$ INFN Roma1, I-00185 Roma, Italy}
\address{$^3$ A. I. Alikhanyan National Science Laboratory, 2 Alikhanyan Brothers Street, 0036 Yerevan, Armenia}
\address{$^4$ Istituto dei Sistemi Complessi, CNR, Via Madonna del Piano 10,
I-50019 Sesto Fiorentino, Italy}
\address{$^5$ SISSA, via Bonomea 265, I-34136 Trieste, Italy}
\address{$^6$ INFN, Sezione di Firenze, I-50019 Sesto Fiorentino, Italy }
\address{$^7$ Department of Physics, University of Trieste, Strada Costiera 11, I-34151 Trieste, Italy} 
\address{$^8$ CNR-IOM DEMOCRITOS Simulation Center, Via Bonomea 265, I-34136 Trieste, Italy}

\ead{\mailto{alessandro.campa@iss.it}, \mailto{v.hovhannisyan@yerphi.am}, \mailto{ruffo@sissa.it} and \mailto{andreatr@sissa.it}}

\begin{abstract}
We derive the phase diagram of the one-dimensional three-state Potts model with an additional mean-field interaction in the canonical ensemble. The free energy is
obtained by mapping the model onto the spin-$1$ Blume-Emery-Griffiths model and solving it by using a Hubbard-Stratonovich transformation combined with the transfer
matrix method. A complex structure with lines of first-order transitions, two triple points and a critical point appears at finite temperature. The phase diagram is
two-dimensional, since there are two adjustable parameters, the nearest-neighbour coupling $K$ and the temperature $T$. We show that the phase diagram does not present
second-order phase transition lines, due to the fact that the order parameter is not a symmetry-breaking one. Quite remarkably, we are able to determine analytically
one of the first-order phase transition lines. We also prove that, when the nearest-neighbour coupling 
$K$ is large and negative, the first-order transition temperature becomes asymptotically independent of $K$.        
\end{abstract}

\maketitle

\section{Introduction}

The interplay between short-range and long-range interactions in physical systems, such as those found in condensed matter and statistical mechanics, has long been
a subject of research \cite{baker63,nagle,kardar1983,chak,lori_ruffo,giacomo_andrea,ourbook}.  To formulate a minimal yet insightful framework for studying this
interplay, one can use a discrete lattice spin model, for instance, a chain. This model would incorporate two types of interactions: nearest-neighbour couplings to
represent short-range forces, and a long-range, mean-field-type component that connects each site to all others.  A variant could be to have couplings that decay as
a power law for large distances \cite{ourbook,fisher} in presence of one or few additional short-range couplings. In the simplest
setting with nearest-neighbour and all-to-all couplings, for classical Ising variables, the one-dimensional (1D) Hamiltonian of the model is typically written as
\begin{equation}
H =  - \frac{J}{2N} \left( \sum_{i} S_i \right)^2 -\frac{K_{NN}}{2} \sum_i S_i S_{i+1} \, ,
\label{NK}
\end{equation}
where $S_i= \pm 1$ denotes Ising spins at lattice sites, $K_{NN}$ is the strength of the short-range (nearest-neighbour) interaction, $J$ quantifies the long-range
mean-field coupling, and $N$ is the total number of sites. This combination can lead, depending on the relative sign of the two couplings, to a competition between a
tendency to local ordering and another one to global coherence, resulting in a rich structure of the phase diagram. Notably, the model exhibits features such as several
types of phase transitions, metastability, and unusual critical behaviour that cannot be captured by models with purely local or purely global interactions. The model
with the Hamiltonian in equation (\ref{NK}), introduced by Nagle \cite{nagle} and Kardar \cite{kardar1983}, serves as a valuable tool in exploring
how non-additive, long-range interactions affect thermodynamic and dynamical properties of interacting systems. An important result is that it exhibits ensemble
inequivalence, with the canonical and microcanonical phase diagrams presenting different phase transition lines \cite{mukamel2005}; see more references
in \cite{campa2019,yang2022}.

A generalization of the Ising model is the $q$-state Potts model, in turn widely studied in statistical mechanics to investigate phase transitions and critical
phenomena \cite{revwu}. In this model, each site $i$ on a lattice is associated with a discrete spin variable $\sigma_i \in \{1, 2, \ldots, q \}$, representing one
of $q$ possible states. The Hamiltonian of the ferromagnetic Potts model is given by
\begin{equation}
\label{pottsgen}
H = -K \sum_{\langle i,j \rangle} \delta_{\sigma_i, \sigma_j},
\end{equation}
where $K>0$ is the nearest-neighbour coupling constant, the sum runs over nearest-neighbour pairs $\langle i, j \rangle$, and $\delta_{\sigma_i, \sigma_j}$ is the
Kronecker delta, equal to $1$ if $\sigma_i = \sigma_j$ and $0$ otherwise. It is clear that the case $q=2$ is equivalent to the Ising model, with just a redefinition
of the coupling energies of a pair of spins that are in the same state or in different states. The Potts model exhibits a rich variety of behaviours depending on the
number of states $q$ and the spatial dimensionality of the system. In two dimensions, the model undergoes a continuous phase transition for $q \leq 4$, while for
$q > 4$ the transition becomes first-order \cite{revwu,baxter_book}. Due to its versatility, the Potts model serves as a theoretical framework for studying not only
magnetic systems, but also problems in percolation, lattice gauge theories, and combinatorics. In Ref. \cite{revwu} the interested reader can find an exhaustive review
dealing with these issues. We remind that in the purely mean-field Potts model, i.e., in the model in which the interaction between $\sigma_i$ and $\sigma_j$ is not
restricted to nearest-neighbour spins, but it is the same for all pairs, the transition is of first-order for $q \ge 3$ \cite{revwu,jsp2005}.

The $q$-state Potts model has a rich critical behaviour, either in the case of only nearest-neighbour coupling or only mean-field interaction. Moreover, it could show
nontrivial effects induced by the competition between short-range and mean-field interactions as in the Nagle-Kardar model (\ref{NK}). In this paper we study a
Potts-type extension of the Nagle-Kardar model. Such a system reveals novel phase transition phenomena, and provides further insight into how nonlocal interactions
influence the collective behaviour in multi-state systems, i.e., in systems where the single components, contrary to the case of Ising spins, can occupy more than two
states. The crucial question is how a nearest-neighbour interaction affects the mean-field term when the number of spin states is larger than two. We focus on the
case $q=3$, since already in this case the mean-field behaviour is different from that of the Ising model; in fact it is well known that the mean-field Ising model has
a second-order phase transition. We consider both cases of positive (ferromagnetic) and negative (antiferromagnetic) nearest-neighbour couplings $K$. Notice that the
Potts model with three states is different from the alternative generalization of the Nagle-Kardar model to the case in which $S$ is a spin $1$ variable (i.e.,
it assumes the values $S=1,0,-1$). This latter model, when there are only all-to-all interactions, has a second-order phase transition (for a discussion of the
spin-$1$ Nagle-Kardar model see Refs. \cite{yang2022,yang2024}).

Summarizing our results, we underline that, in the presence of both the mean-field and the nearest-neighbour interaction the phase transitions remain first-order, and
in the thermodynamic phase diagram (that, as we explain below, is two-dimensional) there are several lines of first-order transitions, with just a single critical point,
that is of a very peculiar nature, being the end-point of three different first-order lines (we devote a section to its detailed study). Two of the first-order lines
extend to infinite absolute values of the nearest-neighbour coupling constant, one to the positive side and the other to the negative side, yet with different
characteristics. In the ferromagnetic case also the transition temperature increases without limits together with the coupling-constant (we are able to find an
analytical expression for this transition line), while in the antiferromagnetic case the transition temperature remains finite, tending to an asymptotic value that
can be analytically obtained.

The structure of the paper is as follows. In section 2 we explicitly give the Hamiltonian of the model and we explain how one can map it to the Hamiltonian of the
spin-$1$ Blume-Emery-Griffiths model \cite{begorig} in order to perform the analytical calculations. In this section we also anticipate that the equilibrium states
never break completely the natural symmetry of the Potts model (in other words, the spontaneous symmetry breaking, always associated to the low temperature behaviour
of the equilibrium states, is only partial), this property allowing to simplify the computations. In section 3 we show our results, presenting the phase diagram in
the canonical ensemble and discussing several examples of the order parameter as a function of the temperature at different values of the nearest-neighbour coupling
constant. Section 3.1 is devoted to the study of the phase diagram at large values of the coupling constant; on the other hand, section 3.2 gives a detailed analysis
of the properties of the critical point. In section 4 we present our conclusions. Two appendices are related to a couple of side issues: in the first we give an a
posteriori justification of the fact that the equilibrium states break only partially the symmetry of the model, while in the second we give an alternative evaluation
of the first-order transition line for large negative values of the nearest-neighbour coupling constant.

\section{The model}
\label{secmodel}

Our model is a 1D three-states Potts model with both a mean-field and a nearest-neighbour interaction, with Hamiltonian given by:
\be
\label{hamil}
H = -\frac{1}{2N}\sum_{i,j} \delta_{S_i,S_j} - K \sum_i \delta_{S_i,S_{i+1}} \equiv H_1 +H_2 \, ,
\ee
where $S_i$ is a spin variable that can assume three different states. There is no need to have a quantitative characterization of the three states other than being
different. However, we are going to map our model onto the mean-field Blume-Emery-Griffiths (BEG) model of spin 1 variables \cite{begorig}, augmented, with respect to
the original model, with a nearest-neighbour interaction, and therefore we characterize the states of $S_i$ with $S_i = \{ -1,0,1\}$. The coefficient $K$ is the
nearest-neighbour coupling constant, and in the coefficient $J/(2N)$ of the mean-field term we can take $J=1$ without loss of generality, by redefining the unit of
energy. As a consequence, the thermodynamic phase diagram is two-dimensional, with coordinates given by the nearest-neighbour coupling constant $K$ and the
temperature $T$.

To perform the mapping to the BEG model we use the following relation, valid for $S_i = \{ -1,0,1\}$:
\be
\label{maprel}
\delta_{S_i,S_j} = \frac{1}{2} S_i S_j +\frac{3}{2} S_i^2 S_j^2 -S_i^2 - S_j^2 +1 \, .
\ee
With this mapping, the mean-field part $H_1$ of the Potts Hamiltonian (\ref{hamil}) becomes
\be
\label{mapmean}
H_1 = -\frac{1}{4N}\sum_{i,j} S_i S_j -\frac{3}{4N}\sum_{i,j} S_i^2 S_j^2 + \sum_i S_i^2 -\frac{N}{2} \, .
\ee
On the other hand, the nearest-neighbour part $H_2$ becomes
\be
\label{mapnear}
H_2 = -\frac{K}{2} \sum_i S_i S_{i+1} -\frac{3K}{2} \sum_i S_i^2 S_{i+1}^2 +2K \sum_i S_i^2 -NK \, .
\ee
Removing constant terms and multiplying by $2$, we then obtain the Hamiltonian
\bea
\label{mapfull}
H &=& -\frac{1}{2N}\left(\sum_i S_i \right)^2 -\frac{3}{2N}\left(\sum_i S_i^2\right)^2 +
\left( 2+ 4 K \right)\sum_i S_i^2 \nonumber \\
&-&K \sum_i S_i S_{i+1} - 3K\sum_i S_i^2 S_{i+1}^2 \, .
\eea
Let us recall the Hamiltonian of the mean-field BEG model, which is
\be
\label{begham}
H = -\frac{1}{2N}\sum_{i,j} S_i S_j -\frac{\widetilde{K}}{2N}\sum_{i,j} S_i^2 S_j^2 + \Delta \sum_i S_i^2 \, ,
\ee
where $\Delta$ and $\widetilde{K}$ are the parameters of the model.
We thus see from Eqs. (\ref{mapfull}) and (\ref{begham}) that the pure mean-field Potts model, i.e., the one with $K=0$, is mapped to the BEG
Hamiltonian with the particular values $\widetilde{K}=3$ and $\Delta =2$, a very peculiar case of the BEG model \cite{pre2017}.
When $K\ne 0$ the Hamiltonian (\ref{mapfull}) becomes the BEG model with this specific parameter choice. However, it is augmented with three terms
proportional to $K$: an additional onsite field and two different nearest-neighbour interactions. The Hamiltonian (\ref{mapfull}) has
two mean-field terms, and its canonical partition function can be computed with the help of the Hubbard-Stratonovich transformation
\be
\label{hubtrans}
\exp (ba^2) = \sqrt{\frac{b}{\pi}} \int_{-\infty}^{+\infty} \dd x \, \exp (-bx^2 + 2abx) \, .
\ee
Performing one such transformation for each one of the two mean-field terms, the partition function is obtained as:
\bea
\label{partfun}
\!\!\!\!\!\!\!\!\!\!\!\!\!\!\!\!\!\!\!Z (\beta,K,N) &\equiv& \sum_{\{S_i\}} \exp \left[ -\beta H \right] = \sum_{\{S_i\}} \exp \left\{
\frac{\beta}{2N}\left(\sum_i S_i \right)^2 +\frac{3\beta}{2N}\left(\sum_i S_i^2\right)^2 \right. \nonumber \\
&& \left. - \left( 2\beta+ 4\beta K \right)\sum_i S_i^2 + \beta K \sum_i S_i S_{i+1} + 3\beta K\sum_i S_i^2 S_{i+1}^2 \right\} \nonumber \\
&=& \frac{\beta \sqrt{3}N}{2\pi}\!\! \int_{-\infty}^{+\infty}\!\!\!\! \dd x \, \int_{-\infty}^{+\infty}\!\!\!\! \dd y \,
\exp \left[-\frac{\beta N}{2}\left(x^2+3y^2 \right) \right] \widetilde{Z}(\beta,K,N,x,y) \, ,
\eea
where
\bea
\label{partfunb}
\widetilde{Z} (\beta,K,N,x,y) &=& \sum_{\{S_i\}} \exp \left\{
\beta x\sum_i S_i + 3\beta y\sum_i S_i^2
- \left( 2\beta+ 4\beta K \right)\sum_i S_i^2 \right. \nonumber \\
&& \left.+ \beta K \sum_i S_i S_{i+1} + 3\beta K\sum_i S_i^2 S_{i+1}^2 \right\} \, .
\eea
The last expression is the partition function of a Hamiltonian with only nearest-neighbour interactions, and it can be computed
using the transfer matrix method. Assuming periodic boundary conditions we can write
\be
\label{partfunc}
\widetilde{Z} (\beta,K,N,x,y) = \sum_{\{S_i\}} \prod_{i=1}^N M_{S_i,S_{i+1}} \, ,
\ee
where $S_{N+1} = S_1$, and where
\bea
\label{partfund}
M_{S_i,S_{i+1}} &=& \exp \left\{ \frac{\beta x}{2}(S_i + S_{i+1}) +\frac{\beta}{2}(3y-2-4K)(S_i^2+S_{i+1}^2) \right. \nonumber \\
&& \left. +\beta K S_i S_{i+1} +3\beta K S_i^2 S_{i+1}^2 \right\} \, .
\eea
Then, in the thermodynamic limit $N\to \infty$ one obtains
\be
\label{partfune}
\widetilde{Z} (\beta,K,N,x,y) = \lambda^N(\beta,K,x,y) \, .
\ee
where $\lambda(\beta,K,x,y)$ is the largest eigenvalue of the $(3\times 3)$ transfer matrix
\be
\label{transmat}
\fl
M(\beta,K,x,y) =  \left(
\begin{array}{ccc}
\er^{ \beta \left( x+3y-2 \right)} &  \er^{ \beta \left( \frac{1}{2}x +\frac{3}{2}y -1-2K \right)} & \er^{ \beta \left(3y-2-2K \right)}  \\
\er^{ \beta \left( \frac{1}{2}x +\frac{3}{2}y -1-2K \right)} & 1 &  \er^{ \beta \left(-\frac{1}{2}x+\frac{3}{2}y -1-2K \right)} \\
\er^{ \beta \left(3y-2-2K \right)} & \er^{ \beta \left(-\frac{1}{2}x+\frac{3}{2}y -1-2K \right)} & \er^{ \beta \left( -x +3y-2 \right)}
\end{array} \right).
\ee
Therefore in the thermodynamic limit we have
\bea
\label{partfunf}
&&Z (\beta,K,N) = \\ &=& \frac{\beta \sqrt{3}N}{2\pi} \int_{-\infty}^{+\infty}\!\! \dd x \, \int_{-\infty}^{+\infty}\!\! \dd y \,
\exp \left\{ N \left[-\frac{\beta}{2}\left(x^2+3y^2 \right) + \ln \lambda(\beta,K,x,y) \right] \right\} \nonumber  \, .
\eea
Then in the same limit the free energy per spin $f(\beta,K)$ is given by
\be
\label{freeen}
\fl
\,\,\,\,\,\,\,\,\beta f(\beta,K) = - \lim_{N\to \infty} \frac{1}{N} \ln Z (\beta,K,N) =
\min_{x,y}\left[ \frac{\beta}{2}(x^2+3y^2) - \ln \lambda(\beta,K,x,y) \right]  \, .
\ee
This expression shows that the free energy, and consequently the thermodynamic phase diagram, is obtained by a minimization problem with respect to
the two auxiliary variables $x$ and $y$. The values $x^*$ and $y^*$ realizing the minimum are, respectively, the values of the magnetization
$m= (\sum_i S_i)/N$ and of the quadrupole moment $q=(\sum_i S_i^2)/N$ in the corresponding equilibrium state \cite{pre2017}.
In the following, we will use $x$ and $y$ when referring to the auxiliary variables with respect to which the optimization has to be performed, and we will
use $m$ and $q$, instead, when referring to the actual values of these variables, i.e., the magnetization and the quadrupole moment, in the equilibrium state. We have
chosen to adhere to the usual convention to use the symbol $q$ for the quadrupole moment. In the literature this symbol is also usually employed to identify a
$q$-state Potts model, with $q$ different spin states (and in fact in the Introduction we have used this symbol while referring to general Potts models), but this
cannot generate confusion, since in this work we consider exclusively a three-state Potts model. We also point out that in the following we will refer to
spin state as one of the states $a$, $b$ and $c$ (or $-1$, $0$ and $+1$ in the BEG representation) in which the spin can be, while we will refer simply to state
when talking about the equilibrium state of the whole system, as e.g., represented by the equilibrium values $m$ and $q$.

The fact that the three spin states are characterized, as remarked above, only by being different spins states, without further quantitative property, has important
consequences, that allow to simplify the optimization problem (\ref{freeen}), arguing as follows. Let us denote as $(n_a,n_b,n_c)$ the occupation fractions of the
three spin states in a state of the  system, being it in or out of equilibrium. Since there is a nearest-neighbour interaction, the three fractions $(n_a,n_b,n_c)$
are not sufficient to fully characterize a state, and also the value of $\sum_i \delta_{S_i,S_{i+1}}$ is necessary, but this is not relevant for our present
argument. Because of the nature of the interactions in the Potts model, any state $(n_a,n_b,n_c)$ is six-fold degenerate if the three fractions are all different; the
other five states are obtained by the permutations of the fractions. On the other hand, the state will be three-fold degenerate if two of the fractions are equal, and
it will be non-degenerate if the fractions are all equal to $1/3$. Restricting now to equilibrium states, the above identification, in the mapping to the BEG model, of
the three spin states $a$, $b$ and $c$ of the Potts Hamiltonian with, respectively, the three spin states $+1$, $0$ and $-1$ of the BEG Hamiltonian, implies
that $m=n_a-n_c$ and $q=n_a+n_c$. It is also necessary to take into account the constraint $n_a+n_b+n_c=1$, so that actually only two of these three variables are
necessary to specify the occupation fractions of the three spin states (coherently, in the BEG representation we need only the two parameters $m$ and $q$); assume that
as these two variables we take the occupation fractions of the spin states $a$ and $b$, i.e., $n_a$ and $n_b$. Suppose now that an equilibrium state is characterized by
the fractions $(n_a^*,n_b^*)$, with $n_a^* \ne n_b^*$ and also $n_c^* = 1-n_a^* -n_b^*$ different from both $n_a^*$ and $n_b^*$. According to what said above concerning
the degeneracy of the states, also the states $(n_b^*,n_a^*)$, $(n_a^*,n_c^*)$, $(n_c^*,n_a^*)$, $(n_b^*,n_c^*)$ and $(n_c^*,n_b^*)$ will be equilibrium states,
i.e., the equilibrium state is six-fold degenerate. On the other hand, if $n_a^*\ne n_b^*=n_c^*$, then the equilibrium state will be only three-fold degenerate, the
other two states being given by $(n_b^*,n_a^*)$ and $(n_b^*,n_b^*)$ (analogous situations occur when $n_b^*\ne n_a^*=n_c^*$ and when $n_c^*\ne n_a^*=n_b^*$). Going now
to the BEG representation, these cases will translate in the following for the values of the magnetization $m$ and the quadrupole moment $q$. When
$n_a^*\ne n_b^* \ne n_c^* \ne n_a^*$ the six degenerate equilibrium states will have $[m,q]$ equal to
$[\pm (n_a^*-n_c^*),n_a^*+n_c^*]$, $[\pm (n_a^* -n_b^*),n_a^*+n_b^*]$ and $[\pm (n_b^*-n_c^*),n_b^*+n_c^*]$. When
$n_a^*\ne n_b^* =n_c^*$, the three degenerate equilibrium states will have $[m,q]$ equal to $[\pm (n_a^*-n_b^*), n_a^*+n_b^*]$
and $[0,2n_b^*]$ (and analogous cases when the two equal fraction are $n_a^*$ and $n_b^*$ or $n_a^*$ and $n_c^*$). Finally,
when $n_a^*=n_b^*=n_c^*=1/3$, the non-degenerate equilibrium state will have $[m,q] = [0,2/3]$.

We remark the relation of the above possibilities of the equilibrium states with the simpler but similar situation that occurs in the case of Ising spins. In fact,
in this case the Hamiltonian is invariant under the spin inversion, but we know that at low temperature the equilibrium state has a magnetization $m$ different from zero.
However, for any such state there is the corresponding degenerate state with opposite magnetization. Then, while in Ising models the spontaneous symmetry breaking
leads to only two degenerate equilibrium states, for the three-state Potts model we can have, in principle, equilibrium states in which the symmetry is totally
broken, and there are six degenerate states, or equilibrium states in which the symmetry is only partially broken, and there are three degenerate states.
The difference between the case of a six-fold degenerate equilibrium state and that of a three-fold degenerate state is that in the former we will find
(restricting ourselves, as allowed by symmetry, to non negative values of the auxiliary variable $x$) three distinct couples $[m,q]$ that satisfy the optimization
problem (\ref{freeen}), and all of them with $m\ne 0$; in the latter we will find one couple $[m,q]$ with $m\ne 0$ and a couple $[0,q]$ (with a different $q$).

The important point is the following. We have found that for all different values of $\beta$ and $K$ that we have tried, we fall, without exceptions, in the second
situation. In other words, the equilibrium state of the three-state Potts model is always one with at least two equal occupation fractions for the three spin states,
so that the symmetry is broken only partially. This was already found for the purely mean-field ($K=0$) three-states Potts model \cite{jsp2005}. This property allows
to study the optimization problem (\ref{freeen}) putting from the start $x$ equal to $0$, then solving an optimization problem only with respect to $y$,
simplifying considerably the numerical effort. It is not difficult to derive that, having found an equilibrium state with
$[0,q]$, the other degenerate state must have $[m_1,q_1]=[\pm (3q/2-1),1-q/2]$. In Appendix A we provide an argument that shows the reason why the equilibrium states
of this Potts model are all with at least two equal occupation fractions.

In conclusion, the free energy per spin is given by
\be
\label{freeenb}
\beta f(\beta,K) = \min_{y}\left[ \frac{3\beta}{2}y^2 - \ln \lambda(\beta,K,y) \right] \, ,
\ee
where $\lambda(\beta,k,y)$ is the largest eigenvalue of the transfer matrix (\ref{transmat}) with $x$ set equal to $0$, that for convenience we rewrite
here
\be
\label{transmat0}
M(\beta,K,y) =  \left(
\begin{array}{ccc}
\er^{ \beta \left( 3y-2 \right)} & \er^{ \beta \left( \frac{3}{2}y -1-2K \right)} & \er^{ \beta \left(3y-2-2K \right)}  \\
\er^{ \beta \left( \frac{3}{2}y -1-2K \right)} & 1 & \er^{ \beta \left( \frac{3}{2}y -1-2K \right)} \\
\er^{ \beta \left(3y-2-2K \right)} & \er^{ \beta \left( \frac{3}{2}y -1-2K \right)} & \er^{ \beta \left( 3y-2 \right)}
\end{array} \right).
\ee
The eigenvalues of this matrix are the solutions of a cubic equation in $\lambda$. It is not difficult to compute that the largest eigenvalue, posing for
convenience of notation $\exp[\beta(3y-2)]\equiv a$ and $\exp[-2\beta K]\equiv b$, is given by
\be
\label{expllam}
\lambda(\beta,K,y) = \frac{1}{2}\left\{ a\left(1+b\right)+1+\sqrt{\left[a\left(1+b\right)-1\right]^2+8ab^2}\right\} \, .
\ee
Therefore, Eq. (\ref{freeenb}) can be rewritten as
\bea
\label{freeenb1}
&&\beta f(\beta,K) = \min_y \widetilde{\phi}(\beta,K,y) \\ &=& \min_{y}\left[ \frac{3\beta}{2}y^2 +\ln 2
- \ln \left\{ a\left(1+b\right)+1+\sqrt{\left[a\left(1+b\right)-1\right]^2+8ab^2}\right\}\right] \, , \nonumber
\eea
where we have defined the function $\widetilde{\phi}(\beta,K,y)$ for later reference.

Some observations are in order. We have remarked above that the value of $y$ that realizes the minimum in the last expression, is equal to the equilibrium
value of the quadrupole moment $q$, a quantity that can take values between $0$ and $1$. Therefore, although the search for the minimum,
from Eq. (\ref{freeen}), in principle is in the whole range of $y$, it is clear that the minimum will occur for a value of $y$ between $0$ and $1$.
We note that the order parameter represented by the quadrupole moment has a different nature with respect to the order parameter represented by the
magnetization $m$, an order parameter that is more commonly associated to the study of the phase diagram of spin systems, and that for our case,
as we have explained, can be chosen to have always the value $0$ in the equilibrium states. In fact, while the Hamiltonian enjoys the invariance with respect
to the operation $m \to -m$, an analogous invariance does not exist for the quadrupole moment (and neither for a more general operation $(q-q_0) \to -(q-q_0)$ with a
reference value $q_0$ different from $0$). This fact is relevant for the structure of the phase diagram of long-range systems: as studied in details in Ref. \cite{bb2005},
these systems, when there is a symmetry like $m \to -m$, behave differently from those without such symmetry. We will come back to this point by commenting the
results that we are going to expose in the next section.

\section{Results}
\label{secresults}
As remarked above, the thermodynamic phase diagram is two-dimensional, with coordinates $(K,T)$, with the temperature $T=1/\beta$ (we choose units in which the Boltzmann
constant $k_B$ is equal to $1$). For any given $K$ and $T$ the equilibrium value of the order parameter $y$ is obtained from the optimization problem (\ref{freeenb1}).
A necessary condition for a value $y$ to be the equilibrium one is the vanishing, for that value, of the derivative of $\widetilde{\phi}(\beta,K,y)$ with respect to $y$;
however, this is not sufficient, since a vanishing derivative could be associated not only to a minimum, but also to a maximum and to an inflexion point. Moreover, there
could be more than one local minimum, and the equilibrium would correspond to the absolute minimum. A peculiarity of our system is related to this issue. In fact, performing
the partial $y$-derivative of $\widetilde{\phi}$, one finds that, regardless of the values of $\beta$ and $K$, at $y=2/3$ this derivative always vanishes. However,
we also find that this value of $y$ corresponds to the global minimum only for sufficiently high temperatures $T$; in other words, for any given $K$, for all temperatures
larger than a value dependent on $K$, the equilibrium value of the quadrupole moment is equal to $2/3$. This is consistent with the fact that at high temperatures we
expect the three spin states to be equally populated, but this result shows that this is not achieved asymptotically with increasing temperature, but already at
finite temperature. As our plots will show, furthermore, the change to an equilibrium value of $q$ equal to $2/3$ occurs through a first-order phase transition.

Before giving the overall results for the phase diagram, we briefly consider what we expect for $T=0$. To this aim, it is convenient to consider the original Hamiltonian
(\ref{hamil}) and determine the configuration with the lowest energy among the configurations with equal number of spins in the states $+1$ and $-1$, since we have
found that we can restrict ourselves to the configurations with $m=0$. For positive $K$ it is clear that the ground state is the one with $q=0$, i.e., with all the
spins in the state $0$. According to the Hamiltonian (\ref{hamil}) its energy per spin is equal to $-(1/2)-K$. For negative $K$, on the other hand, we have to compare this
energy with the one that pertains to the configuration with $q=1$ and with alternating spins in $+1$ and $-1$; its energy per spin is equal to $-(1/4)$. This is smaller than
$-(1/2)-K$ when $K < -(1/4)$. In conclusion, at $T=0$ there is a first-order phase transition at $K=-(1/4)$ between a state with $q=0$ and all spins in the state $0$,
and a state with $q=1$ with alternating spins in the states $+1$ and $-1$.

\begin{figure}[htbp]
\begin{center}
\begin{tabular}{cc}
 \includegraphics[clip, trim=0cm 0cm 0cm 0cm, width=0.46\textwidth]{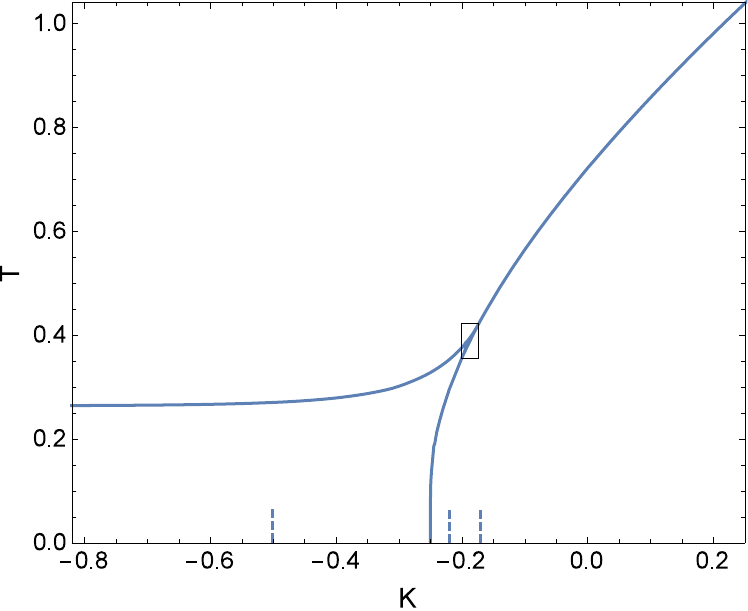} &
 \includegraphics[clip, trim=0cm 0cm 0cm 0cm, width=0.47\textwidth]{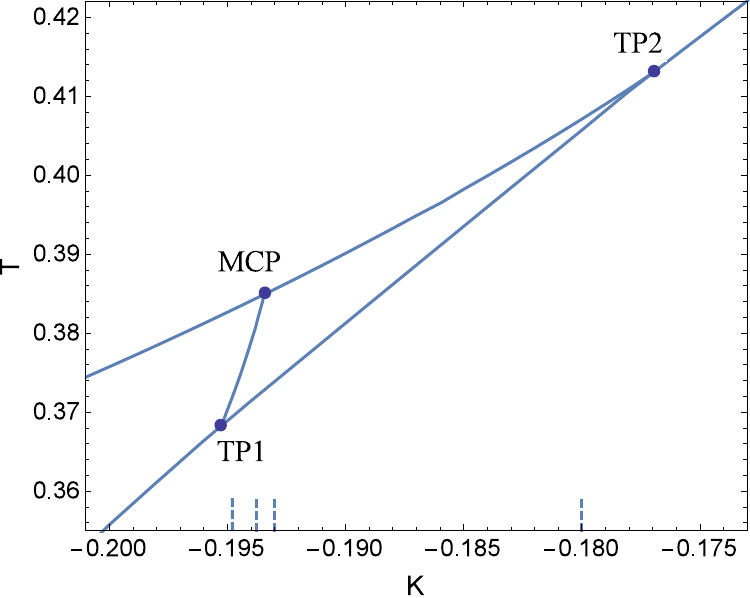}
\end{tabular}
\caption {The $(K,T)$ phase diagram. In the left plot we show the complete phase diagram, while in the right plot we zoom in the region marked with the box in the left plot.
The three points TP1, MCP and TP2 in this plot cannot be conveniently displayed at the scale of the left plot. All lines are first-order transition lines. The points TP1
and TP2 are triple points, with coordinates $(-0.19525,0.36820)$ and $(-0.17681,0.41328)$, respectively. The point MCP, with coordinates $(-0.19345,0.38486)$, is a critical
point. The small dashed lines, on the bottom axis of both plots, mark the $K$ values for which we present, in the following figures, the curves of the quadrupole moment
$q$ as a function of the temperature $T$, showing the first-order transitions. We remark that all three dashed lines on the left plot are at $K$ values outside the range
shown in the right plot, and that the first and second dashed lines on the right plot are for $K$ values between that of TP1 and that of MCP, while the third one is for
a $K$ value larger than that of MCP.}
\label{fig1}
\end{center}
\end{figure}

We now proceed with the results obtained from the minimization problem (\ref{freeenb1}). They are summarized in the $(K,T)$ phase diagram plotted in figure \ref{fig1}.
Just one small warning. Since the expression (\ref{freeenb1}) is based on the Hamiltonian (\ref{mapfull}) obtained by mapping the Potts model onto the BEG model, mapping in
which for convenience, as explained just before that Hamiltonian, we have multiplied all terms by $2$, the values of the temperature at which the various phase transitions
occur are twice those for the original model (\ref{hamil}).

In the left plot of figure \ref{fig1} we show the complete phase diagram, presenting lines of first-order phase transitions, in which the equilibrium value of the
quadrupole moment $q$ undergoes a jump; we show examples of such behaviour below. In the region of the phase diagram which is above all the lines the quadrupole moment
$q$ is equal to $2/3$. The lines meet in the region that in the plot is marked with a small box. That region is blown up and shown
in details in the right plot of the figure. In the left plot we see a line starting at $T=0$, at the value of $K$ equal to $-(1/4)$, obtained above as the value of
first-order transition at vanishing temperature. The asymptotic behaviours of the other lines, the one extending to $K\to -\infty$ and the one extending
to $K\to +\infty$, respectively, are obtained in section \ref{secres1}. We first comment, however, on the behaviour of the phase transition lines in the small region
shown in the right plot. In this plot we see that there is a short line, the one going from the point denoted by TP1 to the point denoted by MCP, which is not visible
at the scale of the left plot. Therefore in that small region there are actually three points, the two just mentioned and the point denoted by TP2, in which phase
transition lines meet. The coordinates of these three points are the following. TP1: $(-0.19525,0.36820)$; MCP: $(-0.19342,0.38490)$; TP2: $(-0.17681,0.41328)$.
Since the transitions are first-order, the points TP1 and TP2 are triple points. At first glance one would say the same for the point MCP; however, as we will prove
in section \ref{secres2}, this point is actually a very peculiar critical point. The fact that we find points that are triple points and a critical point in our phase
diagram is in agreement with the general behaviour of long-range systems analyzed in Ref. \cite{bb2005}: we have a Hamiltonian with one parameter, $K$, so that the
phase diagram is two-dimensional; first-order transitions are associated with codimension $0$ singularities, in turn associated to lines in the phase diagram, while
triple and critical points are associated to codimension $1$ singularities, in turn associated to isolated points in the phase diagram. Also the absence of lines of
second-order transitions can be understood in this framework. In fact, second-order transitions would be associated to codimension $0$ singularities in the case in
which they depend on an order parameter, such as $m$, with respect to which the Hamiltonian is symmetric (e.g., it is invariant for $m \to -m$). But our phase
transitions depend on the order parameter $q$, the quadrupole moment, with respect to which the Hamiltonian does not have the analogous invariance, as already
emphasized. In this case, second-order transitions are associated to codimension $1$ singularities, i.e., to points in the 2D phase diagram.

\begin{figure}[htbp]
\begin{center}
\begin{tabular}{cc}
 \includegraphics[clip, trim=3.7cm 7.7cm 3.7cm 8.5cm,  width=0.47\textwidth]{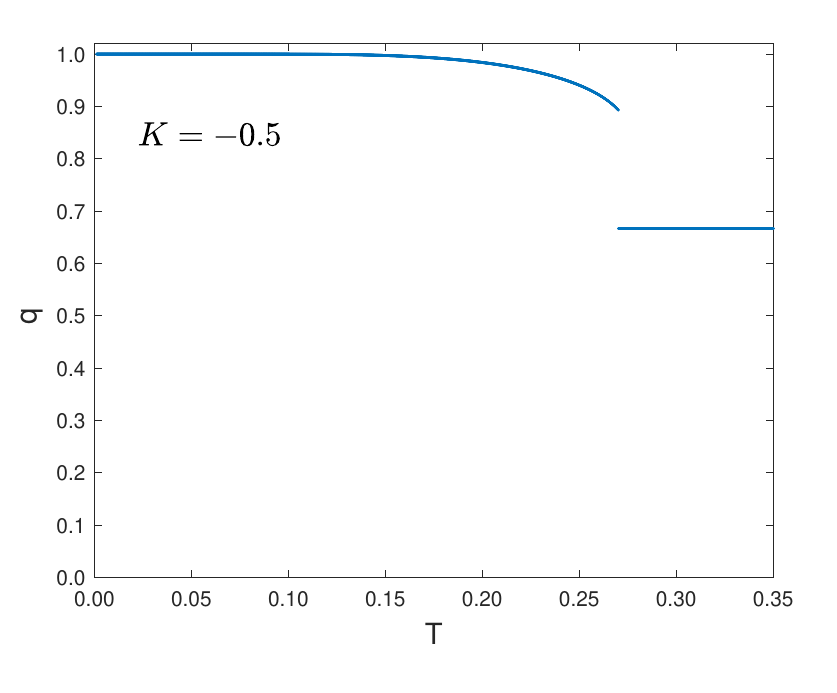} &
 \includegraphics[clip, trim=3.7cm 7.7cm 3.7cm 8.5cm, width=0.47\textwidth]{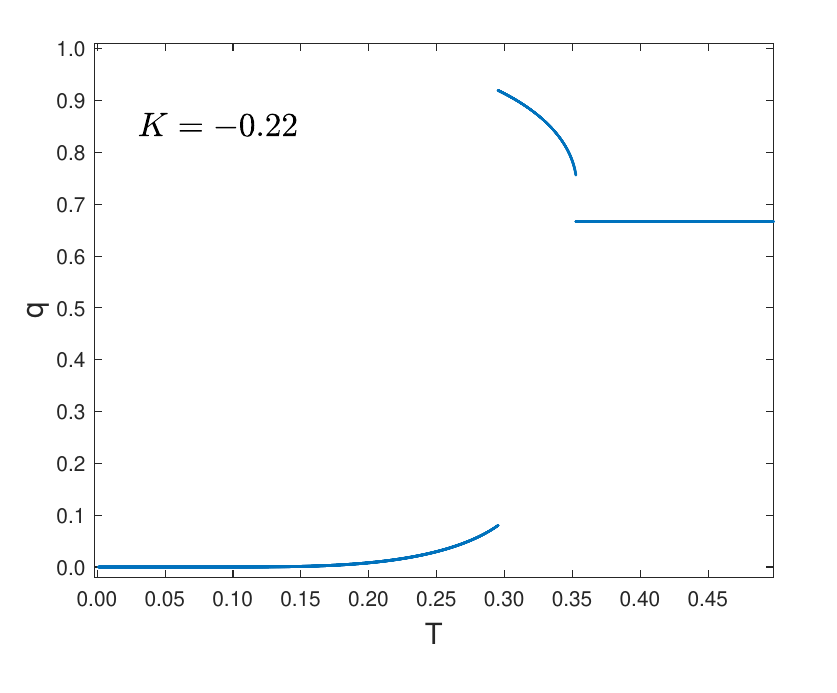}
\end{tabular}
\caption {The quadrupole moment $q$ as a function of the temperature $T$ for a fixed value of $K=-0.5$ (left plot) and $K=-0.22$ (right plot), showing the first-order
transitions with the corresponding jump of $q$. As expected from the phase diagram in figure \ref{fig1}, for $K=-0.5$ there is one phase transition, while there are two
phase transitions for $K=-0.22$. Also, the plots show clearly that for all temperatures larger than that at the phase transition (or that at the last phase transition
if there is more than one), $q$ is constant equal to $\frac{2}{3}$. This occurs for all values of $K$. We finally note that for $T\to 0$ the quadrupole moment tends
to $1$ for $K=-0.5$, while it tends to $0$ for $K=-0.22$, coherently with the fact that the ground state has $q=1$ for $K<-\frac{1}{4}$, while it has $q=0$ for
$K>-\frac{1}{4}$. For a better visualization the plots, as well the left plot in figure \ref{fig3} and the plots in figure \ref{fig4}, have a bottom and a top axis
corresponding to values of $q$ slightly smaller than $0$ and slightly larger than $1$, although these are obviously unphysical values.}
\label{fig2}
\end{center}
\end{figure}

\begin{figure}[htbp]
\begin{center}
\begin{tabular}{cc}
 \includegraphics[clip, trim=3.7cm 7.7cm 3.7cm 8.5cm,  width=0.47\textwidth]{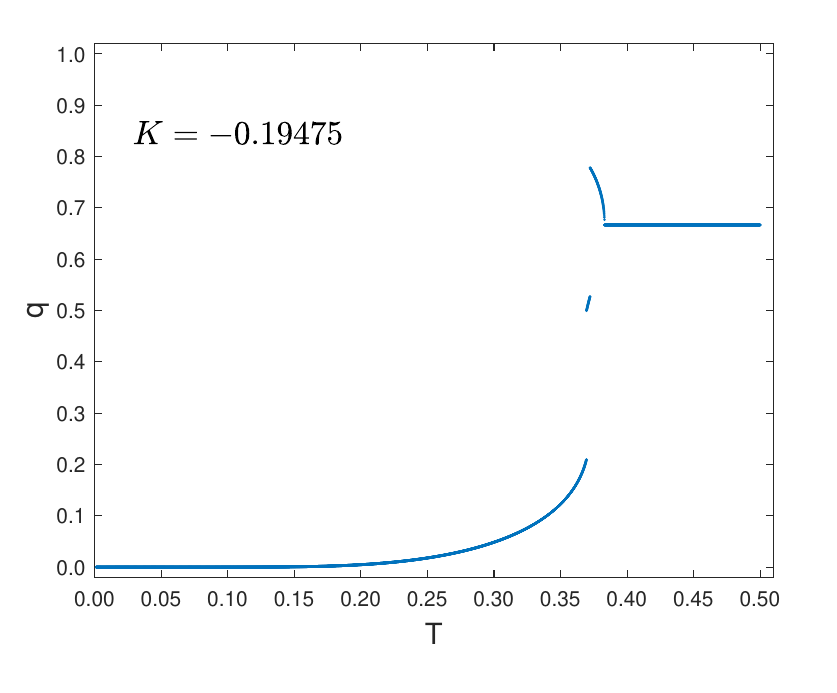} &
 \includegraphics[clip, trim=3.7cm 7.7cm 3.7cm 8.5cm, width=0.47\textwidth]{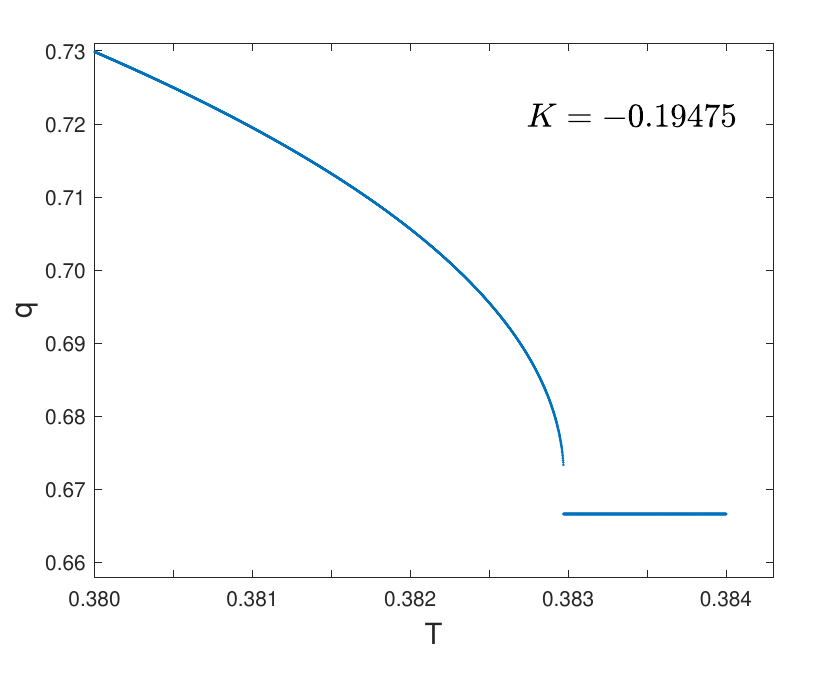}
\end{tabular}
\caption {The quadrupole moment $q$ as a function of the temperature $T$ for a fixed value of $K=-0.19475$. In agreement with the phase diagram, see figure \ref{fig1},
here we have three first-order phase transitions by increasing $T$. In the left plot there is the complete curve, while in the right plot there is a zoom of the region
around the last phase transition by increasing $T$. This plot has also the purpose to show more clearly the (small) jump of $q$ occurring at the transition.}
\label{fig3}
\end{center}
\end{figure}

\begin{figure}[htbp]
\begin{center}
\begin{tabular}{cc}
 \includegraphics[clip, trim=3.7cm 7.7cm 3.7cm 8.5cm,  width=0.47\textwidth]{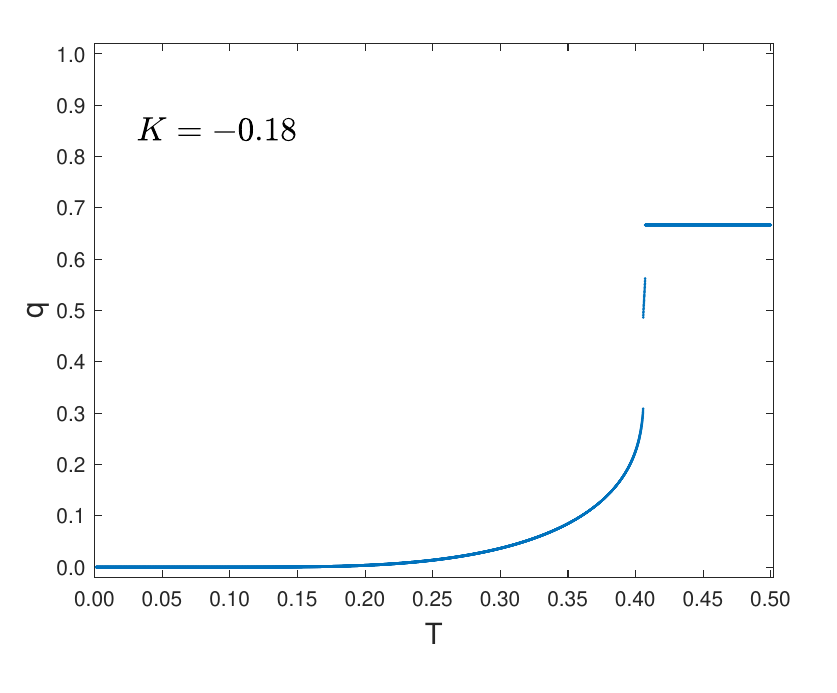} &
 \includegraphics[clip, trim=3.7cm 7.7cm 3.7cm 8.5cm, width=0.47\textwidth]{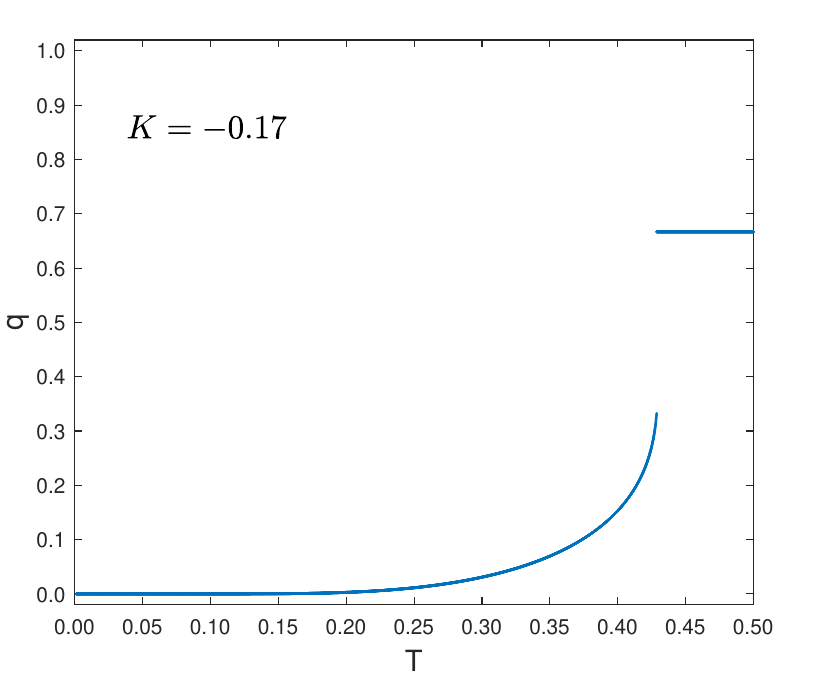}
\end{tabular}
\caption {The quadrupole moment $q$ as a function of the temperature $T$ for fixed values of $K=-0.18$ (left plot) and $K=-0.17$ (right plot). As expected from the phase
diagram in figure \ref{fig1}, for $K=-0.18$ there are two phase transitions, while there is one phase transition for $K=-0.17$.}
\label{fig4}
\end{center}
\end{figure}

We now show some plots of the quadrupole moment $q$ as a function of the temperature $T$ at fixed values of $K$, in order to see the jump of $q$ associated with the
first-order phase transition. In particular, we have chosen for $K$, as examples, the following values: $K=-0.5$, $K=-0.22$, $K=-0.19475$, $K=-0.18$ and $K=-0.17$.
The plots for $K=-0.5$ and $K=-0.22$ are in figure \ref{fig2}; figure \ref{fig3} concerns the value $K=-019475$, while figure \ref{fig4} shows the cases $K=-0.18$
and $K=-0.17$. For the value $K=-0.19475$ we show two plots, the second being the zoom of the curve for a narrow range of $T$, for the reason that we explain shortly.
As one infers from the two plots in figure \ref{fig1}, one can expect that, by increasing $T$, there is one phase transition for $K=-0.5$ and $K=-0.17$, while there
are two phase transitions for $K=-0.22$ and $K=-0.18$, and three phase transitions for $K=-0.19475$. Let us comment these plots.

We first note that, in agreement with the fact that the equilibrium state at $T=0$, as we have determined above, corresponds to $q=1$ for $K< -(1/4)$ and to $q=0$ for
$K> -(1/4)$, the curve for $K=-0.5$ starts at $q=1$ (left plot of figure \ref{fig2}), while all the others start at $q=0$. We also note the following. As we have
anticipated, in all cases, for sufficiently large temperature the equilibrium value of the quadrupole moment is $q=2/3$, and this value is not reached asymptotically,
but it is obtained for all temperatures larger than a $K$-dependent value; moreover, the value $q=2/3$ occurs after a first-order phase transition, as we see in all
plots of figure \ref{fig2}, figure \ref{fig3} and figure \ref{fig4}.

Our next comment concerns the comparison of the curve for $K=-0.22$, in the right plot of figure \ref{fig2}, and the curve for $K=-0.19475$, in figure \ref{fig3}. The
latter value of $K$ is only slightly larger than that of the point TP1, as can be seen in the phase diagram in figure \ref{fig1}. The comparison shows that at the
triple point TP1 there are actually three phases, one with the value of $q$ around $0.2$, one with $q$ around $0.5$, and one with $q$ around $0.8$, and that the phase
with $q$ around $0.5$ is not present for $K$ smaller than the value at TP1. The zoom in the right plot of figure \ref{fig3} has the purpose to show better the jump at
the first-order transition occurring at the largest temperature.  An analogous comparison can be made between the curve for $K=-0.18$ and that for $K=-0.17$, in the
two plots of figure \ref{fig4}. These two values of $K$ are somewhat smaller and somewhat larger, respectively, of the value at the triple point TP2. The comparison
shows that actually at this point there are again three phases, one with $q$ around $0.3$, one with $q$ around $0.55$, and one with $q=2/3$; the phase with $q$ around
$0.55$ is not present for $K$ larger than the value at TP2. As can be guessed from the phase diagram in figure \ref{fig1}, for all values of $K$ larger than that at TP2
the plot of $q$ versus $T$ will be qualitatively similar to that seen at $K=-0.17$, with one phase transition from a value of $q$, smaller than $2/3$, to $q=2/3$. Actually,
we find that, for $K$ larger than that at TP2, at the transition the order parameter $q$ always jumps from $1/3$ to $2/3$. In this $K$ range the value of the temperature
at the transition increases with $K$ in the way determined in section \ref{secres1}.

In section \ref{secres2}, after explaining how the critical point MCP can be determined, we show two further plots, for values of $K$ just smaller and just larger than the
value at MCP, to see the peculiarity of this critical point.

\subsection{The asymptotic behaviour of the transition lines for $K\to \pm \infty$}
\label{secres1}
The asymptotic behaviour of the transition lines for $K\to \pm \infty$ can be obtained as in the following. We begin by considering the case $K\to +\infty$. In this case
we can actually obtain the analytical expression of the whole transition line from the TP2 point to the right, at increasing $K$. For this we note that the numerical
computations have shown that the first-order transition occurring at this line is always associated to a jump form $q=1/3$ to $q=2/3$. This has suggested to investigate
the behaviour of the function $\widetilde{\phi}(\beta,K,y)$ at a particular $K$-dependent value of $\beta$, a value such that the function is symmetric with respect to
the value $y=1/2$. If we find such a value, then, since we have already noticed that the function has a vanishing derivative at $y=2/3$ for any value of $\beta$ and $K$,
then for that particular value of $\beta$ the function will have also a vanishing derivative for $y=1/3$. This is a necessary, although not sufficient, condition to
have a first-order transition between $y=1/3$ and $y=2/3$. We proceed as follows. Remembering that $a=\exp[\beta(3y-2)]$, after a very simple manipulation we first
rewrite the function $\widetilde{\phi}(\beta,K,y)$, defined in Eq. (\ref{freeenb1}), as
\bea
\label{phinew}
&&\widetilde{\phi}(\beta,K,y) = \beta \left( \frac{3}{2}y^2 -\frac{3}{2}y +1 \right) \\
&& + \ln 2 - \ln \left\{ a^{\frac{1}{2}}\left(1+b\right)+a^{-\frac{1}{2}}+\sqrt{\left[a^{\frac{1}{2}}\left(1+b\right)-a^{-\frac{1}{2}}\right]^2+8b^2}\right\} \, .
\nonumber
\eea
For the two particular values $y=0$ and $y=1$ we thus have
\bea
\label{phipart}
\fl
\widetilde{\phi}(\beta,K,0) &=& \beta +\ln 2 -\ln \left\{\er^{-\beta}\left(1+b\right)+\er^{\beta}
+\sqrt{\left[\er^{-\beta}\left(1+b\right)-\er^{\beta}\right]^2+8b^2}\right\}  \, , \\
\fl
\widetilde{\phi}(\beta,K,1) &=& \beta +\ln 2 -\ln \left\{\er^{\frac{\beta}{2}}\left(1+b\right)+\er^{-\frac{\beta}{2}}
+\sqrt{\left[\er^{\frac{\beta}{2}}\left(1+b\right)-\er^{-\frac{\beta}{2}}\right]^2+8b^2}\right\} \nonumber  \, .
\eea
Equating these two expression we get either $\er^{\beta}=1$ or $\er^{\beta}=\left(1+b\right)^2$. The first solution means $T=\infty$ and does not interest us;
the second is the solution we seek. Denoting with $\beta^*$ this $K$-dependent temperature, and writing the function $\widetilde{\phi}$ for this particular temperature,
we obtain, after posing $y=z+\frac{1}{2}$,
\bea
\label{phinewp}
\fl
&&\widetilde{\phi}(\beta^*,K,z) = \beta \left( \frac{3}{2}z^2 +\frac{5}{8} \right) \\
\fl
&& + \ln 2 - \ln \left\{ \left(1+b\right)^{3z+\frac{1}{2}}+\left(1+b\right)^{-3z+\frac{1}{2}}
+\sqrt{\left[\left(1+b\right)^{3z+\frac{1}{2}}-\left(1+b\right)^{-3z+\frac{1}{2}}\right]^2+8b^2}\right\} \, , \nonumber
\nonumber
\eea
which is manifestly even in $z$. We have thus obtained that the phase transition occurs at the temperature $T$ such that
$\er^{\frac{\beta}{2}} = (1+b)$. Remembering that $b=\exp[-2\beta K]$, this relation can be rewritten as
\be
\label{firstorderex}
K = -\frac{T}{2} \ln \left( e^{\frac{1}{2T}} -1 \right) \, ,
\ee
giving the analytical expression of the first-order transition line to the right of the point TP2. For values of $K$ smaller than that at TP2, the $K$-dependent value
of $\beta$ such that the function $\widetilde{\phi}$ is even with respect to $y=\frac{1}{2}$ is still defined, and the values $y=\frac{1}{3}$ and $y=\frac{2}{3}$
still correspond to points where the value of $\widetilde{\phi}$ is the same and its first derivative with respect to $y$ vanishes; however, they are not global minima
in $y$ of $\widetilde{\phi}$. The asymptotic behaviour of (\ref{firstorderex}) for large $K$ and $T$ is $K=\frac{T}{2}\ln(2T)$.

We now consider the transition line that extends towards $K\to -\infty$. The numerical results suggest that this line tends to an asymptotic value, and that at the
transition the quadrupole moment $q$ jumps from a value that is asymptotically the same to the value $q=\frac{2}{3}$. This is supported by
the following analysis. When $K$ is negative and very large in absolute value, and $\beta$ is finite, $b$ becomes very large, and we can approximate the expression
$a\left(1+b\right)\pm 1$ with just $ab$, so that the function $\widetilde{\phi}(\beta,K,y)$ in Eq. (\ref{freeenb1}) is approximated by
\be
\label{phiapprox}
\widetilde{\phi}(\beta,K,y) =  \frac{3\beta}{2}y^2 +\ln 2 + 2\beta K - \ln \left\{ a+\sqrt{a^2+8a}\right\} \, .
\ee
We see that the parameter $K$ now appears just as an additional term, and the minimization with respect to $y$ does not involve it. This proves that for large
negative $K$ the phase transition line tends to an asymptotic value of the temperature, that can be computed as $T\approx 0.26354$, as found in the numerical results,
and that at the transition the quadrupole moment jumps from a value that asymptotically is given by $q\approx 0.90053$ to the value $q=2/3$. In Appendix B we provide
an interesting alternative evaluation of this result, based on the direct counting of the number of microscopic states.

\subsection{The peculiar critical point MCP}
\label{secres2}
Here we show that the point marked with MCP in the right panel of figure \ref{fig1} is a critical point, but of a very peculiar nature. In a system in which there is
no symmetry with respect to the order parameter, as in our case, the critical point is found by equating to zero the first three derivatives, with respect to the order
parameter, of the function to be extremized, the function $\widetilde{\phi}$ in our case. Since this function depends on three quantities, $\beta$, $K$ and $y$, it is
clear that, as anticipated above, this procedure actually determines an isolated point (or in principle isolated points) in the $(K,T)$ phase diagram. To proceed we note
that the numerical results suggest that the critical point is associated to a continuous transition occurring at the peculiar value of the quadrupole moment order
parameter equal to $\frac{2}{3}$. Then, instead of computing the $y$-derivatives of the function $\widetilde{\phi}$ we perform, equivalently, its power expansion around
that value of $y$, knowing already that the first order term will vanish, and we will determine the particular values of $\beta$ and $K$ for which also the second and
the third order terms vanish. To this purpose in the function $\widetilde{\phi}$ defined in Eq. (\ref{freeenb1}) we pose $y=\frac{2}{3} + \epsilon$ and we perform a
power expansion up to the third order in $\epsilon$. After a straightforward computation we arrive at
\be
\label{expphi}
\fl
\widetilde{\phi}(\beta,K,y) = \frac{2}{3}\beta + \ln 2 - \ln (2+4b) + c_2(\beta,K) \epsilon^2 + c_3(\beta,K) \epsilon^3 + O(\epsilon^4) \, ,
\ee
where the coefficients $c_2$ and $c_3$ are given by
\bea
\label{coeffexp}
c_2(\beta,K) &=& \frac{3}{2}\beta - \frac{\beta^2}{3b} (b +2)  \\
\label{coeffexp3}
c_3(\beta,K) &=& -\frac{\beta^3}{9b^2} ( b^2 -2b -2) \, .
\eea
Equating to zero the coefficients $c_3$ we get a value for the parameter $b$. We recall that $b$ is defined by
$b=\exp[-2\beta K]$. The expression in brackets in Eq. (\ref{coeffexp3}) has one negative and one positive root, and only the last one is acceptable; it is equal to
$b=1+\sqrt{3}$. Plugging this value in the expression of $c_2$ and equating it to zero we get the
value of $\beta$ and thus of the temperature $T$ at the critical point; we find $T=\frac{2}{9}\sqrt{3} \approx 0.38490$. Inserting this value in the expression for $b$
we get the value of the coupling parameter $K$ at the critical point, $K= -\frac{\sqrt{3}}{9}\ln \left( 1+\sqrt{3}\right) \approx -0.19342$.

The peculiarity of this critical point resides in the following. Generally a critical point is the end point of a first-order transition line. However we see in
figure \ref{fig1} that there is no end point of such lines (apart the end of one of the lines at $T=0$). The coordinates that we have just provided are those of the point
marked with MCP in the figure. What happens is that three first-order transition lines converge to that point, as we note in the figure, one from the left, one from
the right and one from below. Each one of them marks a first-order transition associated to a jump in the quadrupole moment $q$, and this jump, approaching the
point $MCP$, tends to vanish, according to what occurs at a critical point, and for all three lines the value of $q$ at the critical point approaches $\frac{2}{3}$.

\begin{figure}[htbp]
\begin{center}
\begin{tabular}{cc}
 \includegraphics[clip, trim=3.7cm 7.7cm 3.7cm 8.5cm,  width=0.47\textwidth]{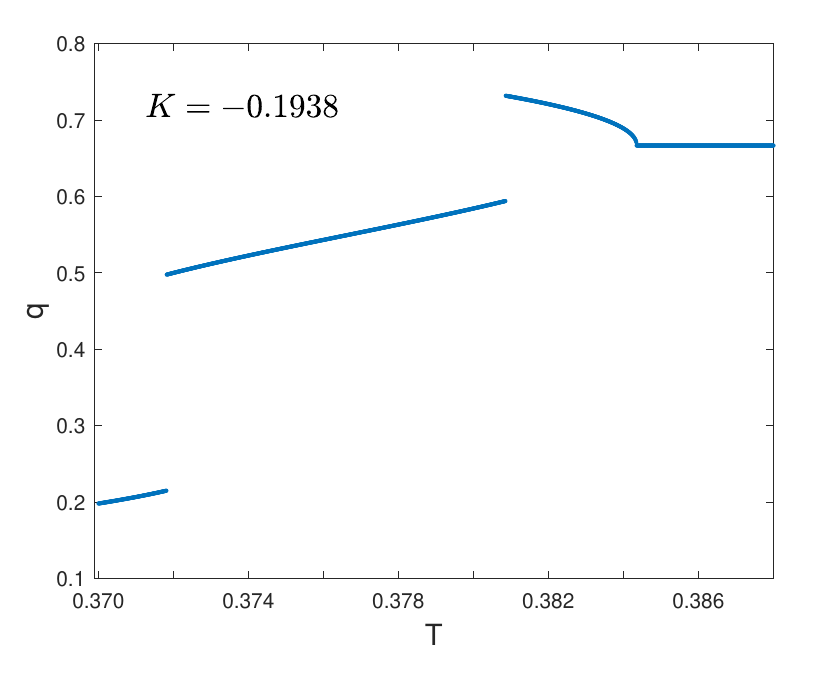} &
 \includegraphics[clip, trim=3.7cm 7.7cm 3.7cm 8.5cm, width=0.47\textwidth]{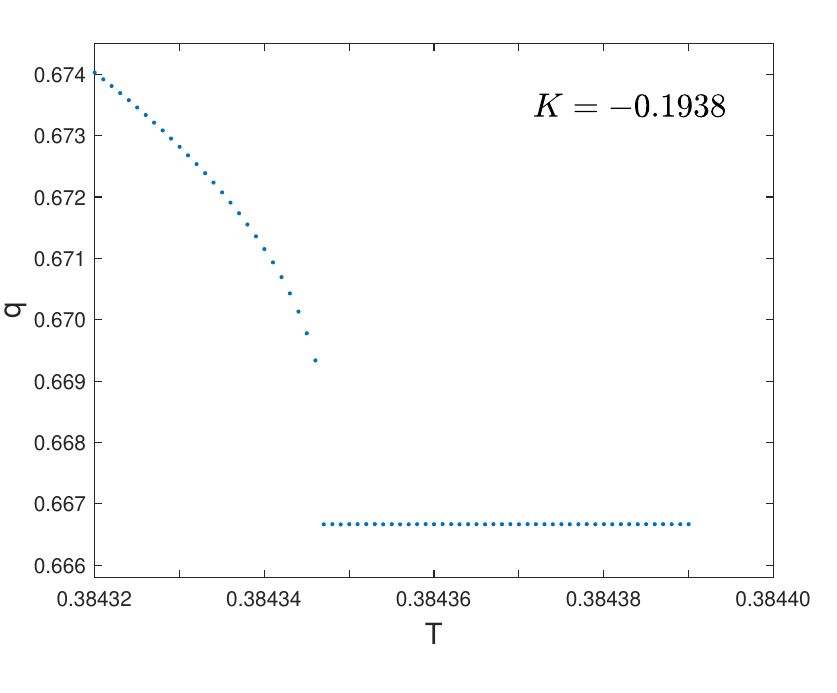}
\end{tabular}
\caption {The quadrupole moment $q$ as a function of the temperature $T$ for a fixed value of $K=-0.1938$, which is slightly smaller than the value at the point MCP.
Then, according to the phase diagram in figure \ref{fig1}, we have three phase transitions. The right plot is a zoom of a quite narrow range in $q$ and $T$ (such that
the space between the dots corresponding to the computed values is now visible), that shows more clearly that also the last transition is first-order, although with a
very small jump in $q$.}
\label{fig5}
\end{center}
\end{figure}

\begin{figure}[htbp]
\begin{center}
\begin{tabular}{cc}
 \includegraphics[clip, trim=3.7cm 7.7cm 3.7cm 8.5cm,  width=0.47\textwidth]{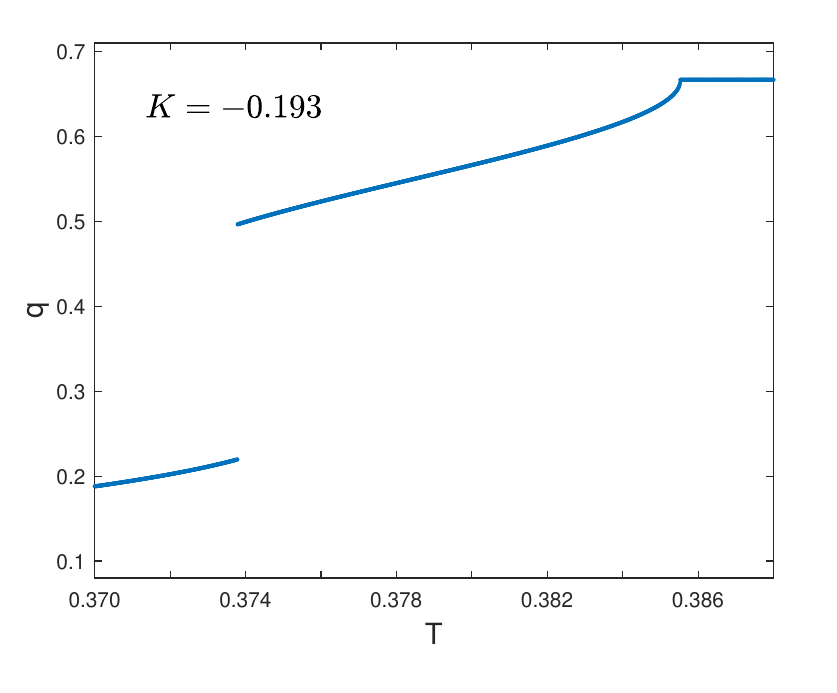} &
 \includegraphics[clip, trim=3.7cm 7.7cm 3.7cm 8.5cm, width=0.47\textwidth]{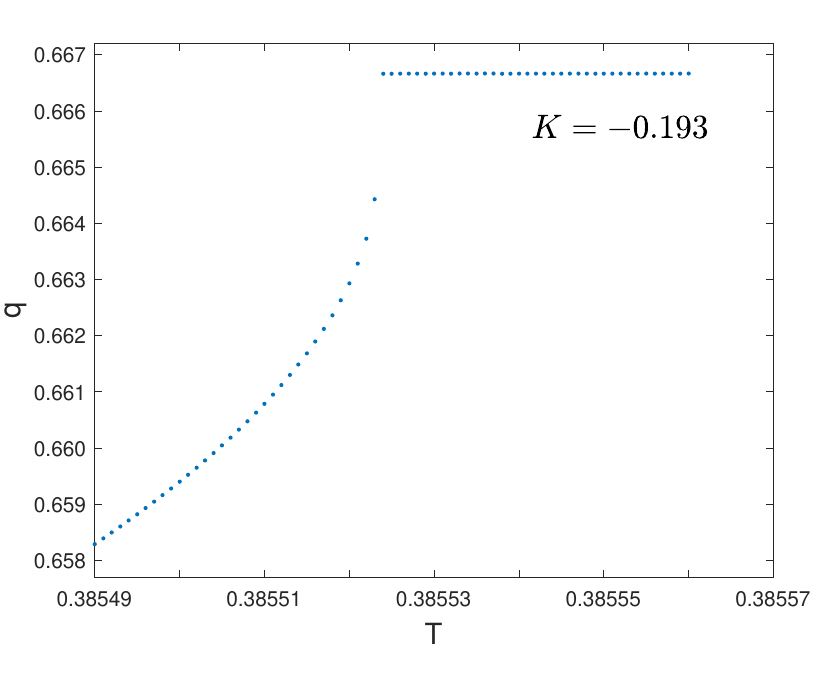}
\end{tabular}
\caption {The quadrupole moment $q$ as a function of the temperature $T$ for a fixed value of $K=-0.193$, which is slightly larger than the value at the point MCP. Then,
according to the phase diagram in figure \ref{fig1}, we have two phase transitions. The right plot is a zoom of a quite narrow range in $q$ and $T$ (such that the space
between the dots corresponding to the computed values is now visible), that shows more clearly that also the last transition is first-order, although with a very
small jump in $q$.}
\label{fig6}
\end{center}
\end{figure}

We conclude by showing two further plots of $q$ versus the temperature $T$ for two values of $K$, one just smaller and one just larger than the value at the critical
point MCP. In figure \ref{fig5} we have the case $K=-0.1938$, and in figure \ref{fig6} the case $K=-0.193$. As above for the case $K=-0.19475$, we show two plots for
each case, the plot in the right part of the figure being a zoom for a narrow range of $T$. As before, the purpose of the zoomed plots is to show that actually we have
a first-order transition, although the jump in the value of $q$ is so small that it is not evident in the plot at the larger scale on the left.

\section{Discussion and conclusions}

We have studied the phase diagram of a 1D three-states Potts model with both a mean-field and a nearest-neighbour interaction. As it occurs for spin models, the
simultaneous presence of these two kinds of interactions causes the occurrence of a complex and interesting thermodynamic phase diagram, with the relevant features
showing up when there is a competition between them. In particular, with a ferromagnetic mean-field interaction we can find these features when the nearest-neighbour
interaction is antiferromagnetic.

We have found useful and interesting to map the three-states Potts Hamiltonian to an extended version of the BEG model, a spin-1 model with mean-field interactions and
an on-site field, obtained by adding $2$ nearest-neighbour interactions. The mapping is based on the relation (\ref{maprel}). The nature of
the Potts model allowed us to study the phase diagram by restricting to equilibrium states of the BEG model with zero magnetization. The phase transition, then, will
involve the order parameter $q$. As explained in section \ref{secmodel}, any such equilibrium state is actually three-fold degenerate, with the exception of the state with
$q=\frac{2}{3}$, which is non-degenerate. This also means, as we have remarked, that the equilibrium states break only partially the symmetry of the Hamiltonian, while in
principle one could have expected a complete symmetry breaking.

Since the Hamiltonian is not symmetric with respect to $q$, continuous transitions correspond, in the 2D thermodynamic phase diagram, to isolated points; they are
associated to codimension-1 singularities \cite{bb2005}. As a consequence, the phase transition lines in the phase diagram, associated to codimension-0 singularities,
correspond to first-order phase transitions. It is interesting also to note that these lines extend to both $K\to +\infty$ and $K\to -\infty$. It is especially remarkable
that for sufficiently large negative $K$ the equilibrium state does not depend (asymptotically) on $K$, and so also the temperature of the first-order transition does not
depend on $K$. This is due to the nature of the interaction of the Potts model, in which, contrary to the usual spin-1 models, the interaction between two spins
distinguishes only two situations, i.e., equal or different spin states.

We could find an analytical expression for the first-order phase transition line for values of $K$ larger than those at the
triple point TP2. It is also remarkable the fact that in this part of the phase transition line, the transitions always occurs
between the values $q=1/3$ and $q=2/3$. While the latter value is explained by the fact that for large enough temperatures this
must be the equilibrium state (similarly to what happens in Ising models, where at high enough temperatures we must have $m=0$),
we cannot envisage an intuitive physical explanation for the former value being always $1/3$.

We have considered the canonical phase diagram. The presence of first-order phase transition lines allows us to make the following comment. We remind that in systems
with long-range interactions the presence of first-order transitions in the canonical ensemble implies ensemble inequivalence: the phase diagram in the microcanonical
ensemble will then be different \cite{ourbook}. The simpler case with $K=0$, i.e., the model with only the mean-field interaction, was analyzed with the large deviation
technique in Ref. \cite{jsp2005} in order to study ensemble inequivalence. According to our results, in the canonical case the order parameter $q$ shows a first-order
transition jumping from the value $1/3$ to the value $2/3$ at the temperature $T=1/(2 \ln(2))\approx 0.72135$ (the solution of equation (\ref{firstorderex}) for $K=0$).
Restating the results of Ref. \cite{jsp2005} in our notations, while the canonical solution confirmed this result, it was found that in the microcanonical ensemble, on
the contrary, there is no phase transition, and the order parameter $q$ goes continuously and monotonically from $q=0$ at the lowest energy (corresponding to $T=0$)
to $q=2/3$ at the highest energy. Furthermore, there is a region of negative specific heat starting from the energy value for which $q\approx 0.415$ up to the highest
energy. Even more marked differences between the ensembles are expected in the general model with $K\ne 0$.

Finally, we mention two interesting avenues for future study. First, it would be highly interesting to investigate the model considered here by varying the spin states
beyond the current number of three, to check if our results could be extended. In the Introduction we have already reminded that in the purely mean-field
Potts model with $q$ states the transition is first-order for $q\ge 3$ (only in this paragraph, as it is clear from the context, $q$ refers to the number of states
of a Potts model, and not to the quadrupole moment). Here we limit to remark that the equilibrium states in these mean-field models are characterized
by having $q-1$ spin states equally populated  \cite{revwu}. In this paper we have seen that this particular feature holds, for $q=3$, also for the model with both a
mean-field interaction and a nearest-neighbour interaction. This may suggest that this holds also for $q>3$. In other words, the property that the equilibrium states break only partially the symmetry of the Hamiltonian, could be present also for the general $q$-state Potts model with mean-field and nearest-neighbour interactions.
The study of this possibility could in principle be performed by extending our analysis that uses both a Hubbard-Stratonovich transformation and a transfer matrix method.
A map of the Hamiltonian to a spin model, similarly to what has been done in this work, would be useful to put in evidence the partial symmetry breaking, if present.
It can be argued that this map would require a spin model with $q$ states and a considerably more involved representation of the Kronecker delta.
As a second possible direction of investigation, it would be valuable to study the finite-size properties of the model, specifically
in connection with determining the Casimir force, thereby generalizing the results presented in Ref. \cite{dantchev}.

\section*{Acknowledgments}
VH acknowledges the receipt of the grant in the frame of the research project No. SCS 21AG-1C006.
SR acknowledges support from the MUR PRIN2022 project BECQuMB Grant No. 20222BHC9Z.

\appendix

\section{The $m=0$ property of the equilibrium states}

We have found that the equilibrium states of our three-states Potts model have always two equal occupation fractions of the
spin states, this translating, in the BEG representation, to equilibrium states with $m=0$. In this Appendix we provide an
{\it a posteriori} justification of this property. An equilibrium state with $m=0$ means that the occupation fractions $n_a$
and $n_c$ are equal (for simplicity of notation here we do not use the asterisk to denote equilibrium fractions). Let us
focus on the case in which the third occupation fractions, $n_b = 1-2n_a=1-2n_c$, is different from $n_a$. In other words,
let us consider equilibrium states different from those that occur, for any $K$, for all temperatures larger than a
$K$ dependent temperature, and for which all fractions are equal to $1/3$. One could ask why it is always the case that
an equilibrium state with $n_a=n_c\ne n_b$ has a smaller free energy than a state in which, e.g., some of the spin in the
spin state $a$ go to the spin state $c$, resulting in an equilibrium state with the same $n_b$ but with $n_c>n_a$. To
justify this fact, we use the following argument.

In an equilibrium state that in the BEG representation has a given $q$, there are $Nq/2$ spins in the spin state $a$ and $Nq/2$
spins in the spin state $c$. Consider the Potts Hamiltonian restricted to these $Nq$ spins

\be
\label{reshamil}
H = -\frac{1}{2N}\sum_{i,j} \delta_{S_i,S_j} - K \sum_i \delta_{S_i,S_{i+1}} \, ,
\ee
where now for the spin states we consider only $a$ and $c$. Identifying, similarly to the BEG representation,
$a$ with $S=1$ and $c$ with $S=-1$, and using, when $S$ takes only these two values,
$\delta_{S_i,S_j} = \frac{1}{2}(S_i S_j +1)$, the last Hamiltonian is transformed, neglecting constant terms, in
\be
\label{resbhamil}
\fl
H = -\frac{1}{4N}\sum_{i,j} S_iS_j - \frac{K}{2} \sum_i S_iS_{i+1}
= \frac{q}{2} \left[-\frac{1}{2Nq}\sum_{i,j} S_iS_j - \frac{1}{2}\frac{2K}{q} \sum_i S_iS_{i+1} \right] \, .
\ee
The last expression shows that the Hamiltonian is equal to $q/2$ times the Hamiltonian of the Ising model
of a 1D chain of $M$ spins with a mean-field interaction and nearest-neighbour interaction, i.e.,
\be
\label{reschamil}
H = -\frac{1}{2M}\sum_{i,j} S_iS_j - \frac{\widetilde{K}}{2} \sum_i S_iS_{i+1} \, ,
\ee
with $M=Nq$ and $\widetilde{K}=(2K)/q$; see in fact Eq. (\ref{NK}). The canonical $(T,\widetilde{K})$ phase diagram of this model is known \cite{mukamel2005}:
it has a phase transition line that starts at $(T,\widetilde{K})=(0,-0.5)$ and that goes, as a first-order line, up to the tricritical point
$(T,\widetilde{K})=(1/\sqrt{3},(\ln \sqrt{3})/\sqrt{3})\approx (0.57735, -0.31714)$. After the tricritical point,
the transition line continues indefinitely for increasing $T$ and $K$, as a second-order line, its expression
given implicitly by $T= \er^{(K/T)}$. Above the transition line the equilibrium state has zero magnetization, i.e., the occupation
fractions of the two spins states are both equal to $1/2$. If we now denote with $\widetilde{T}(\widetilde{K})$ the phase
transition line of the above Ising model, then the analogous phase transition line of the Hamiltonian in Eq. (\ref{resbhamil})
will be given, writing explicitly also the dependence on $q$, by $T(K,q) = \frac{q}{2}\widetilde{T}(\frac{2K}{q})$.

We can compare this result with the various plots of $q$ vs. $T$ that we have obtained for different values of $K$, in particular
for the values of $T$ smaller than that of the transition to the final high temperature equilibrium state with all occupation
fractions equal to $1/3$ (and thus $q=2/3$). To take into account that in our BEG representation we have multiplied by $2$ the original Hamiltonian, we have to
actually consider the curve $T(K,q)= q\widetilde{T}(\frac{2K}{q})$.
We denote with $q(K,T)$ the inverse (for a given $K$) of this expression.
If this curve $q(K,T)$ for the given $K$ is above the curve $q$ vs. $T$ found for the equilibrium state, this justifies the
fact that the occupation fractions in $a$ and $c$ are equal. In fact, the value $q(K,T)$ for given $K$ and $T$ is the minimal
value of $q$ for which the Ising Hamiltonian (\ref{resbhamil}) has an equilibrium state with $m\ne 0$, i.e., with $n_a\ne n_c$.

\begin{figure}[htbp]
\begin{center}
\begin{tabular}{cc}
  \includegraphics[scale=0.34]{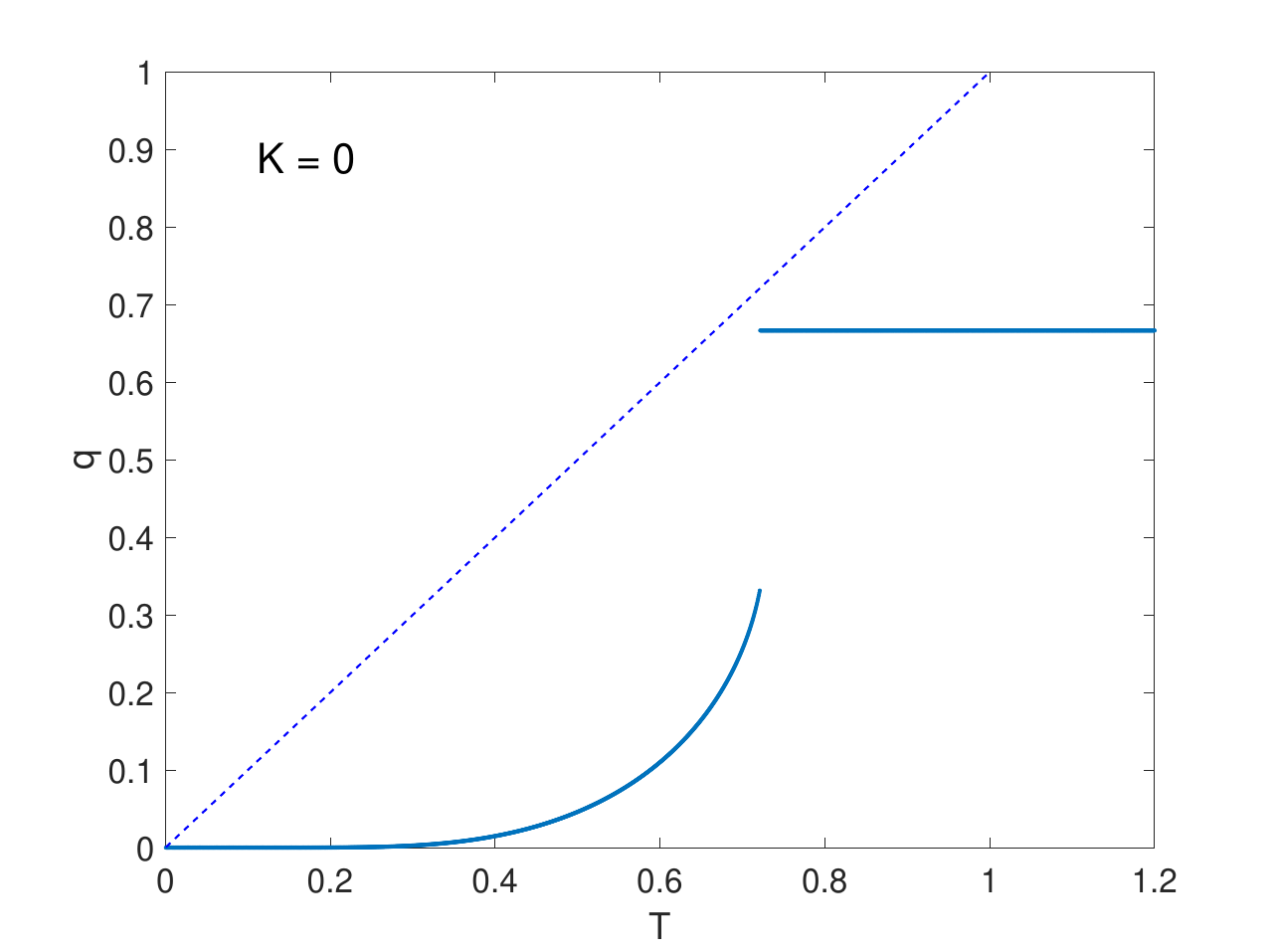} &
\includegraphics[scale=0.34]{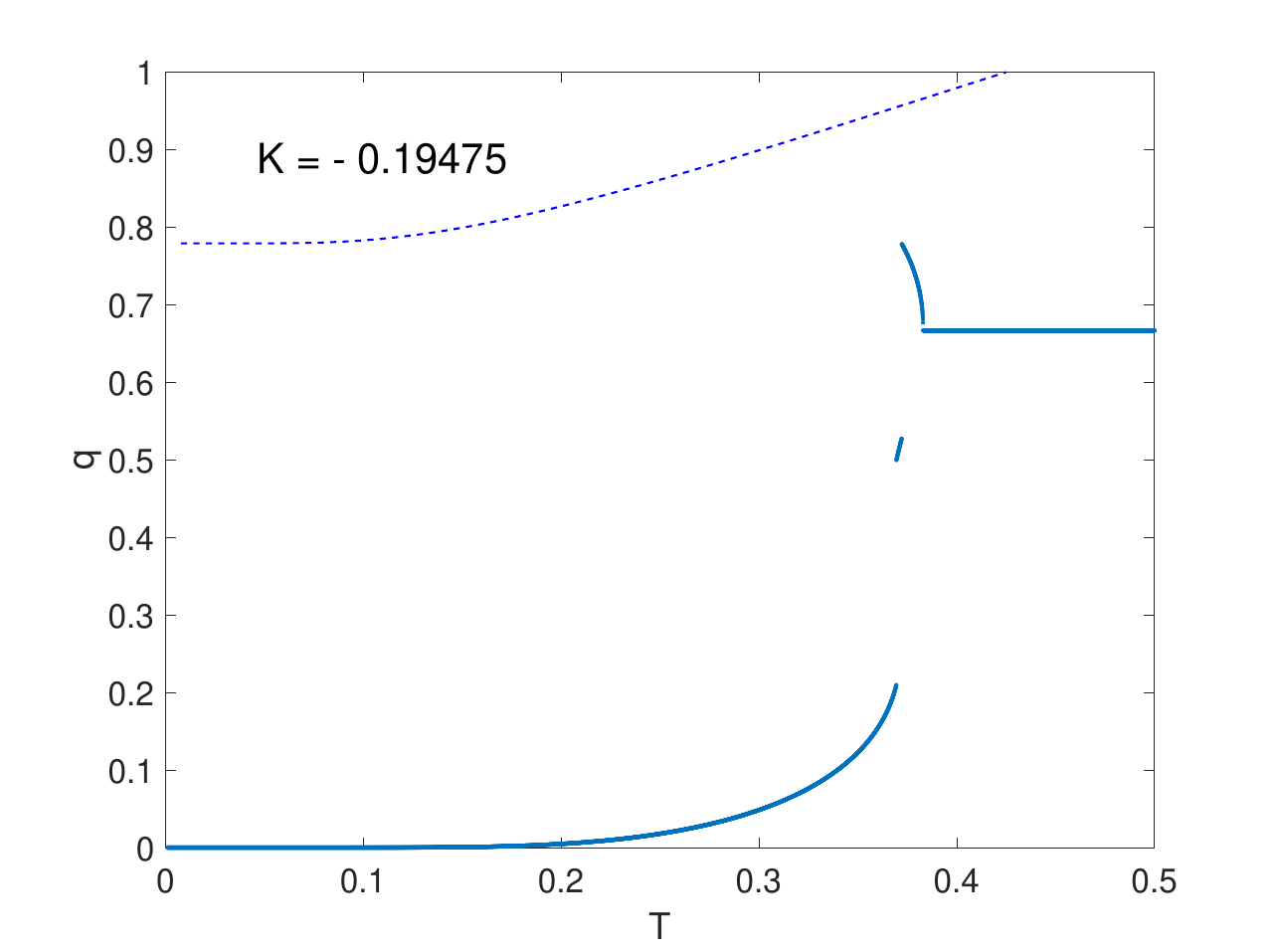}
\end{tabular}
\caption {Full lines: The quadrupole moment $q$ as a function of the temperature $T$ for fixed values of $K$, i.e., $K=0$ in the left panel and $K=-0.19475$ in the
right panel; the latter is the same plot as in the left panel of figure \ref{fig3}. Dashed lines:
the function $q(K,T)$ giving the minimal value of $q$ for which the Ising Hamiltonian (\ref{resbhamil}) has an equilibrium state with $m\ne 0$. Since
the function $\widetilde{T}(x)$ is equal to $1$ for $x=0$, for $K=0$ the dashed line in the left panel is simply $q=T$.}
\label{fig7}
\end{center}
\end{figure}

We have found that this is always the case. We provide in figure \ref{fig7} two examples, one with $K=0$ and another for one of the cases
considered in the main text, $K=-0.19475$. We can see that the curve $q(K,T)$ is above the curve giving the value of $q$ in the
equilibrium state.

\section{Direct counting approach for the case of large negative $K$}

An alternative approach to the case of large negative $K$ is given in this appendix, in which we perform a direct counting of the number of possible states (more
precisely, the leading order of this number in the thermodynamic limit), for a given value of the quadrupole moment and zero magnetization. Using the same notation of
the main text, we will denote with $y$ the quadrupole moment before optimization. Then, for a given
$y$ and $m=0$ we compute the number of possible spin configurations, and let us call this number $\Omega(y,N)$; the logarithm of this number is the entropy corresponding
to the value of $y$. The configurations are all those that, while satisfying these constraints, are also such that no pair of nearest-neighbour spins are in the same
state; in fact, for large negative $K$ two nearest-neighbour spins cannot be in the same state, otherwise they would contribute a diverging positive value to the energy.
Furthermore, we know that for a given $y$ and $m=0$ and no nearest-neighbour spins in the same state, the energy of the system is equal to (since there are a
fraction $y/2$ of spins in one state, a fraction $y/2$ in another state, and a fraction $1-y$ in the third state) $-N[ (y/2)^2 + (y/2)^2 + (1-y)^2]=-N[ 1 -2 y +(3/2)y^2]$
(we remind that in the BEG mapping we have multiplied the Hamiltonian by $2$, therefore the factor $1/2$ is absent in the expression of the energy). Let us call this
energy $E(y,N)$. Calling $\widetilde{f}(T,y)= \lim_{N\to \infty} (1/N) [E(y,N) - T \ln \Omega(y,N)]$, then the free energy per particle will be given
by $f(T) = \min_y \widetilde{f}(T,y)$. We now proceed to the computation of $\Omega(y,N)$. Without the constraint of no nearest-neighbour spins in the same state the
computation would be trivial; on the other hand, the computation of the energy, that would now depend also on $K$, would be much more difficult, and for this reason in
the general case the easiest procedure is the one using the Hubbard-Stratonovich transformation. On the other hand, with the mentioned constraint
the expression of the energy is trivially obtained, and the computation of $\Omega(y,N)$ can be performed, although not so easily as in the case without the constraint.
We propose two ways of computing $\Omega(y,N)$, the first using an actual counting, and the second employing an indirect, but perhaps more elegant, counting.

\subsection{The actual counting of the number of states}

We say that between the $i$-th and the $(i+1)$-th spin there is a kink if the states of the two spins are different. For what has been said above, all configurations we
have to consider are those in which for every pair of spin there is a kink. With periodic boundary conditions (in any case, the boundary conditions are not important
in the thermodynamic limit) all our configurations have $N$ kinks. If there was no $y$ value and no $m$ value to satisfy, $\Omega(y,N)$ would trivially be equal to $2^N$,
since any spin (e.g., going from left to right in the chain) could be placed in any one of the two states different from that of its left neighbour. But we have to
satisfy $y$ and $m$. If $N_{+1}$, $N_0$ and $N_{-1}$ are the number of spins in, respectively, the state $+1$, $0$ and $-1$, and $Q=Ny=N_{+1}+N_{-1}$ is the total
quadrupole moment, then $N_0=N-Q$, and from $m=0$ we have $N_{-1}=N_{+1}=Q/2$. Because of the constraint that no nearest-neighbour spins can be in the same state, it
must be $N/2\le Q \le N$ (we can assume $N$ even, since its parity is not relevant in the thermodynamic limit). In fact, if $Q$ is smaller than $N/2$, it is not
possible to place $N-Q$ spins in the state $0$ without having no two of them in nearest-neighbour spins.  We now denote with $L_{a,b}$ the number of kinks in which the
left spin in a pair is in state $a$ and the right spin is in state $b$; $a$ and $b$ can be $(+1,0,-1)$, but it must be $a\ne b$, so there are six variables $L_{a,b}$.
They are not all independent. Actually, for given $N$ and $y$ one can express five of them as a function of the sixth. Let us take $L_{-1,0}$ as the independent quantity,
and assume that it has the value $R$ (in a moment we will give the range in which $R$ can be chosen). Without showing the easy evaluation, we just mention that the other
quantities are then equal to: $L_{-1,+1}=Q/2-R$, $L_{+1,0}=N-Q-R$, $L_{+1,-1}=3Q/2+R-N$, $L_{0,-1}=N-Q-R$, $L_{0,+1}=R$. The quantity $R$ must satisfy
$\max (0,N-3Q/2) \le R \le \min (Q/2,N-Q)$. Let us now consider the spins in the state $-1$. The spins to the right of each one of them must be either in the
state $+1$ or in the state $0$. Therefore $L_{-1,0}+L_{-1,+1}=Q/2$. This means, using the known expression of the binomial coefficients, that there are
\be
{\cal N}_1 = \frac{\left( \frac{Q}{2}\right)!}{R! \left( \frac{Q}{2}-R\right)!} \, ,
\ee
different ways to put the $L_{-1,0}$ kinks starting from $R$ of the $Q/2$ spins in the state $-1$. Analogously, we can choose the $L_{0,+1}$ kinks in
\be
{\cal N}_2 = \frac{\left( N-Q \right)!}{R! \left( N-Q-R\right)!}   \, ,
\ee
different ways, and the $L_{+1,0}$ kinks in
\be
{\cal N}_3 = \frac{\left( \frac{Q}{2} \right)!}{\left( N-Q-R\right)! \left( \frac{3Q}{2} -N+R\right)!}   \, ,
\ee
different ways. Once one of the kink configurations for each one of the three cases has been chosen, then also the spin configuration of the whole chain is determined
(actually, there would be three different chain configurations for each kink configuration, but this is irrelevant in the thermodynamic limit, since eventually we have
to take the limit of the logarithm of the number of states divided by $N$). Thus, for a given $R$, we count
\be
{\cal N}_{{\rm tot}} = {\cal N}_1 {\cal N}_2 {\cal N}_3
\ee
number of states. Using the Stirling approximation to compute the factorials, and posing $R=Nr$, we arrive at a number of states equal to
\be
\frac{y^{Ny} \left(1-y\right)^{N(1-y)}2^{N(y-1)}}{r^{2Nr}\left(y-2r\right)^{N(y/2-r)}\left(1-y-r\right)^{2N(1-y-r)}\left( 3y-2+2r\right)^{N(3y/2-1+r)}}  \, ,
\ee
The total number of states for a given $y$ would be the sum of this number for all possible values of $r$ in the allowed range. However, as usual in this computations,
we have to consider the leading term in the thermodynamic limit, given by the maximum of this number when $r$ is varied in the allowed range. Taking the derivative of
the logarithm of the last expression, we easily compute that this maximum occurs for $r=(1-y)/2$. Plugging this value in the above expression and taking the logarithm
we arrive at
\be
\label{omega1}
\fl
\ln \Omega (y,N)= N \left[ y\ln y - (1-y)\ln(1-y) +(1-y)\ln 2 -(2y-1)\ln (2y-1)\right]  \, ,
\ee
where we remind that $1/2\le y \le 1$. Thus, at the end we have
\bea
\label{count1}
f(T) &=& \min_y \left\{ -\left(1-2y +(3/2)y^2\right) \right.\\
&-& \left.T\left[ y\ln y - (1-y)\ln(1-y) +(1-y)\ln 2 -(2y-1)\ln (2y-1) \right] \right\}  \, . \nonumber
\eea
This optimization problem gives exactly the same result of Eq. (\ref{phiapprox}). Actually, it is not difficult to obtain that in the allowed $y$ range the expression
in (\ref{phiapprox}) and that in (\ref{count1}) have both a vanishing derivative when the relation
$2y=(1+a/\sqrt{a^2+8a})$ is satisfied, with, we remind, $a$ defined soon before Eq. (\ref{expllam}). When there is more than a value of $y$ that satisfies this relation
(and, accordingly to what has been explained above, it is always satisfied for $y=2/3$), one has to choose the one that is a global minimum, and it can be seen that this
value is the same for (\ref{phiapprox}) and (\ref{count1}).

\subsection{A Markov process evaluation}

We can arrive at the same expression (\ref{omega1}), and then the same free energy (\ref{count1}), using another approach. Let us suppose that the probabilities of
the $i$-th spin of the chain to be in the states $-1$, $0$ and $-1$ are $P_i(+1)$, $P_i(0)$ and $P_i(-1)$, respectively, and also that, like in a Markov process, there
are transition probabilities from the state of the $i$-th spin to the state of the $(i+1)$-th spin, transitions probabilities which are independent from $i$. Clearly,
the transition probabilities from one state to the same state are zero, since no nearest-neighbour spins must be in the same state. In view of the fact that at the end
the fraction of spins in the $+1$ state and the fraction of those in the $-1$ state must be equal, we are allowed to have the following equations describing our
Markov process
\bea
P_{i+1}(+1) &=& \frac{1}{2}P_i(0) + \gamma P_i(-1) \nonumber \\
P_{i+1}(-1) &=& \frac{1}{2}P_i(0) + \gamma P_i(+1) \nonumber \\
P_{i+1}(0) &=& (1-\gamma)\left( P_i(+1) + P_i(-1) \right)  \, ,
\eea
where $\gamma$ is a number between $0$ and $1$ to be determined. We do this by imposing to be in the stationary situations, in which $P_i$ does not depend on $i$ and
where $P(+1)=P(-1)$. Then, from any one of the three relations above we obtain
\be
\gamma = \frac{2P(+1)-P(0)}{2P(+1)}  \, .
\ee
If we now impose to have a quadrupole moment $q$, we must have
\be
\frac{P(+1)}{P(0)} = \frac{y}{2(1-y)} \, ,
\ee
so that
\be
\gamma = \frac{2y-1}{y} \, .
\ee
Therefore we have a Markov process in which from the state $0$ we have equal probabilities $1/2$ to go either to $+1$ or $-1$, while from the state $+1$ ($-1$) we have
probability $\gamma$ to go to $-1$ ($+1$) and probability $1-\gamma$ to go to $0$. In the first case we have a local entropy equal to $\ln 2$, therefore an effective
number of possible states equal to $2$, while in the second case we have a local entropy equal to $-\gamma \ln \gamma - (1-\gamma)\ln(1-\gamma)
= -\ln [\gamma^{\gamma} (1-\gamma)^{1-\gamma}]$, corresponding to an effective number of states equal to
\be
\left[ \gamma^{\gamma} (1-\gamma)^{1-\gamma}\right]^{-1} = \left( \frac{y}{2y-1}\right)^{\frac{2y-1}{y}}\left(\frac{y}{1-y}\right)^{\frac{1-y}{y}} \, .
\ee
In conclusion, we have an effective number of states equal to $2$ for a fraction $(1-y)$ of the spins, those in the state $0$, and an effective number of states equal to
the above expression for a fraction $y$ of spins, those either in $+1$ or in $-1$. Therefore our number of states $\Omega(y,N)$ is
\bea
\Omega(y,N) &=& 2^{N(1-y)} \left[ \left( \frac{y}{2y-1}\right)^{\frac{2y-1}{y}}\left(\frac{y}{1-y}\right)^{\frac{1-y}{y}} \right]^{Ny} \nonumber \\
&=& \frac{2^{N(1-y)}y^{Ny}}{(2y-1)^{N(2y-1)}(1-y)^{N(1-y)}} \, .
\eea
Taking the logarithm we have exactly the expression (\ref{omega1}).

\end{document}